\documentclass[aps,prl, showpacs,groupedaddress,twocolumn,floatfix,longbibliography]{revtex4-1}

\usepackage{amsmath}
\usepackage{graphicx}
\usepackage[usenames,dvipsnames]{color}
\pdfoutput=1

\usepackage{color}
\usepackage{url}
\usepackage[colorlinks]{hyperref}
\hypersetup{%
        plainpages=true,
        breaklinks=true,
        hypertexnames=false,
        pageanchor=true,
        colorlinks=true,
        linkcolor={blue},
        citecolor={blue},
        urlcolor={black}, %Change to blue when doi has been added to all refs, not just a few
        anchorcolor={black}
      }

\newcommand{\be}{\begin{equation}}
\newcommand{\ee}{\end{equation}}
\newcommand{\bea}{\begin{eqnarray}}
\newcommand{\eea}{\end{eqnarray}}

\renewcommand{\eqref}[1]{\mbox{Eq.~(\ref{#1})}}
\newcommand{\figref}[1]{\mbox{Fig.~\ref{#1}}}

\begin{document}

%\preprint{}

%\title{Generation of Attractive and Repulsive Photons by the Dynamical Casimir Effect in Arrays of Coupled Cavities}
\title{Cavity QED Beyond Rotating Wave Approximation:
	\\Photon Bunching  from the Emission of Individual Qubits}

%Can individual qubits emit photon pairs?

%\title{Quantum control and nonclassical state generation in dynamical Casimir arrays}

\author{Luigi Garziano$^{1}$}
\author{Alessandro Ridolfo$^2$}
\author{Simone De Liberato$^1$} \email{S.De-Liberato@soton.ac.uk}
\author{Salvatore Savasta$^{2}$}

\affiliation{$^1$School of Physics and Astronomy, University of Southampton, Southampton, SO17 1BJ, United Kingdom}
\affiliation{$^2$Dipartimento di Scienze Matematiche e Informatiche, Scienze Fisiche e Scienze della Terra (MIFT), Universit\`{a} di Messina, I-98166 Messina, Italy,}

%\thanks{}

%\date{\today}

\begin{abstract}

Photon antibunching in the light scattered by single quantum emitters is one of the hallmarks of quantum optics, providing an unequivocal demonstration of the quantized nature of the electromagnetic field. Antibunching can be intuitively understood by the need for a two-level system lying in its lower state after emitting a photon to be re-excited into the upper one before a second emission can take place.
Here we show that such a picture breaks down in the ultrastrong light-matter coupling regime, when the coupling strength becomes comparable to the bare emitter frequency. Specialising to the cases of both a natural and a superconducting artificial atom, we thus show that a single emitter coupled to a photonic resonator can emit bunched light. The result presented herein is a clear evidence of how the ultrastrong coupling regime is able to change the nature of individual atoms. 
\end{abstract}

%\keywords{}

\maketitle

%-------------------------------------------------

%======================================
%\section{Introduction}
%=======================================

The nonclassical phenomenon of photon antibunching \cite{Carmichael1976}
was first observed in the resonance fluorescence of sodium
atoms in a low-density atomic beam \cite{Kimble1977}. Since then, this phenomenon has been observed in a variety of single quantum emitters as  trapped ions,  dye molecules \cite{Basche1992},  semiconductor quantum dots \cite{Michler2000, Press2007,Yuan2002}, nitrogen-vacancy center in diamond \cite{Brouri2000,Kurtsiefer2000}, single carbon nanotubes \cite{Hoegele2008}, and superconducting qubits \cite{Houck2007}.
Apart from its fundamental importance, it can be used for the realization of triggered single-photon sources \cite{Michler2000a}, an important building block of quantum technology architectures \cite{Bennett1992, Bouwmeester2000,Knill2001}.
The efficiency of triggered single photon sources can be significantly improved by coupling the quantum emitter to a single mode of an electromagnetic cavity with dimensions comparable to the emission wavelength \cite{Pelton2002, Chang2006,Press2007}.
If the interaction rate $ \lambda $ between the atomic dipole and the electromagnetic field amounts to a non negligible fraction of the atomic transition frequency $\omega_a$, the routinely invoked rotating-wave approximation (RWA) is no longer applicable and the antiresonant terms in the interaction Hamiltonian
significantly change the standard cavity QED scenarios \cite{Liberato2007,DeLiberato2009,Ashhab2010,Ai2010,Cao2010,Cao2011,Stassi2013,Garziano2013,Huang2014,Cacciola2014,Garziano2014,DeLiberato2014,Garziano2015a,Zhao2015,Baust2016,Wang2016,Stassi2016,Garziano2016,Rabl2016,Shen2016}. 
In particular, it has been shown that this regime can significantly modify the statistics of cavity photons \cite{Ridolfo2012,Ridolfo2013,LeBoite2016}. This light-matter ultrastrong coupling (USC) regime has been experimentally reached in a variety of solid state systems \cite{Gunter2009,Forn-Diaz2010,Niemczyk2010,Schwartz2011,Geiser2012,Scalari2012,Gambino2014,Goryachev2014,Maissen2014,Semba2017}. Specifically, superconducting circuits have proven to be the most exquisite platform for microwave on-chip quantum-optics experiments in the USC regime. First- and second-order correlation function measurements have been performed in these systems by using quadrature amplitude detectors and linear amplifiers \cite{Bozyigit:2011aa}. Moreover, the deep strong coupling regime, where the coupling strength becomes comparable or even larger than the atomic and cavity frequencies, has been recently achieved in a superconducting flux qubit tunably coupled to an LC oscillator via Josephson junctions \cite{Semba2017}.

\begin{figure}[!ht]
	\centering
	\includegraphics[scale=0.8]{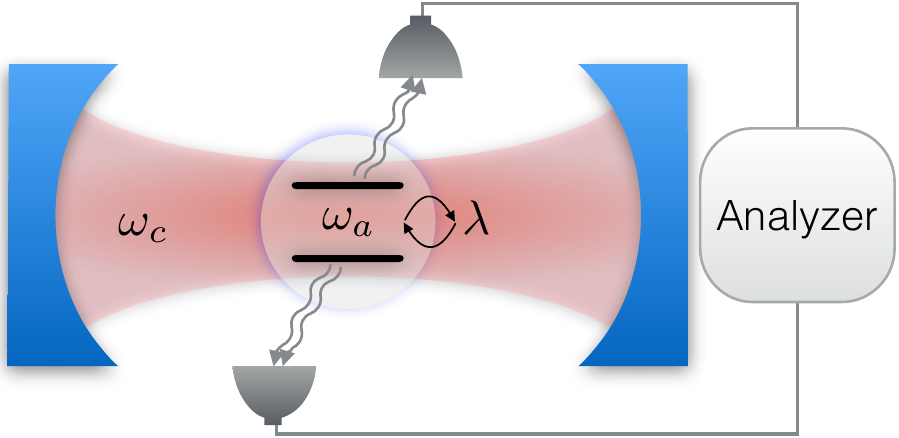}
	\caption{Sketch of the system. A two-level atom is coupled to a resonator mode. Light emitted by the atom into the external modes  is detected by two photodetectors and analyzed to measure two-photon coincidence counting rates.
		\label{fig:1}}
\end{figure}

In this Letter we study the statistics of photons emitted by a two-level system ultrastrongly coupled to a photonic resonator (see \figref{fig:1}). We will discover how, accordingly to the specificity of the system and to the strength of the light-matter coupling, situations arise where the standard antibunching effect does not occur and the two-level system emits bunched light.

The quantum operator describing the electric field can be written as \cite{Milonnibook} 
\begin{equation}\label{Es}
\hat {\bf E} ({\bf r}, t) = \hat {\bf E}_{\rm in} ({\bf r}, t)-
\boldsymbol{\Psi({\bf r})}\,  \ddot{\hat{\sigma}}_x( \tilde t)\, ,
\end{equation}
where $\hat {\bf E}_{\rm in} ({\bf r}, t)$ is the incoming field, $\hat \sigma_x = \hat \sigma_+ + \hat \sigma_-$ with $\hat \sigma_{\pm}$ the atomic rising and lowering operators, $\tilde t = t-r/c$ with $c$ the speed of light, and $\boldsymbol{\Psi({\bf r})} = [{\bf d} - ({\bf d} \cdot {\bf r}) {\bf r}/r^2]/({c^2 r})$  describes the electric field emitted in the far field region by a point dipole with moment $ \bf d $.
According to the quantum theory of photodetection \cite{Glauber1963}, the probability rate for a photon polarized in the $j$ direction to be absorbed by an ideal photodetector at point ${\bf r}$ at time $t$ is proportional to the normal-order correlation function
$\langle \hat{E}_j^- ({\bf r}, t) \hat{E}_j^+ ({\bf r}, t) \rangle$,
where $\hat{E}_j^+$ and $\hat{E}_j^- = (\hat{E}_j^+)^\dag$ are the $j$th Cartesian components of the positive- and negative-frequency electric-field operators. This result can be generalized to coincidence probabilities. For example the probability to detect a photon at point ${\bf r}_1$ at time $t_1$ and a photon at point ${\bf r}_2$ at time $t_2$ is proportional to the normal-order photon-photon correlation function $G^{(2)}({\bf r}_1,t_1;{\bf r}_2,t_2)=\langle \hat E_j^- ({\bf r}_1, t_1) \hat E_j^- ({\bf r}_2, t_2) \hat E_j^+ ({\bf r}_2, t_2)  \hat E_j^+ ({\bf r}_1, t_1) \rangle$. 

When the atom evolves freely at its unperturbed frequency $\omega_a$, we have
\be\label{freeatom}
\hat \sigma_- (t) = \hat \sigma_- (0) e^{-i \omega_a t}\, ,
\ee
and from Eq.~(\ref{Es}) the positive-frequency field takes the form
\begin{equation}\label{Es+}
\hat  {\bf E}^+ ({\bf r}, t) = \hat  {\bf E}_{\rm in}^+ ({\bf r}, t) + \boldsymbol{ \Psi}({\bf r})\, \omega_a^2\,  \hat \sigma_- ( \tilde t).
\end{equation}
According to Eq.~(\ref{Es+}), the normal-ordered, zero-delay second-order correlation function $G^{(2)}({\bf r}, t;{\bf r}, t)  = \langle \hat E_j^- ({\bf r}, t) \hat E_j^- ({\bf r}, t) \hat E_j^+ ({\bf r}, t)  \hat E_j^+ ({\bf r}, t) \rangle$, proportional to $\langle   \hat \sigma_+  \hat \sigma_+ \hat \sigma_- \hat \sigma_- \rangle$, vanishes  since $\hat \sigma_- \hat \sigma_- = 0$. Indeed, only a delayed emission of the second photon is possible, so that $G^{(2)}({\bf r}, t;{\bf r}, t) < G^{(2)}({\bf r}, t;{\bf r}, t + \tau)$ for $\tau >0$.

The same result holds when the quantum interaction with the free-space electromagnetic field is taken into account, having care to redefine the atomic transition frequency to include the Lamb-shift. In the case in which the electromagnetic field interacting with the atom is significantly affected by a photonic structure as an optical cavity, Eq.~(\ref{Es+}) can still be safely used if the light-matter system is in the so-called weak coupling regime. In this regime, where the atom-field coupling rate is smaller than the decay rates of both the field mode and the atomic excitation, the presence of the cavity only induces a modification of the spontaneous emission rate and of the scattered field $\boldsymbol{ \Psi}({\bf r})$. 
In the opposite case, when the interaction between the atomic dipole and the electromagnetic field enters the strong coupling regime, the coupling changes the energy eigenstates and Eq.~(\ref{freeatom}) does not apply anymore. Nevertheless, since the shifts in the atomic frequency are of the same order of $\lambda$, in the strong coupling regime, when $\lambda/\omega_{\rm a}\ll1$, they are small enough so that Eq.~(\ref{Es+}) is still a good approximation. The situation changes in the USC regime, when $\lambda$ amounts to a non-negligible fraction of $\omega_a$ and the time evolution of $\hat{\sigma}_-$ can differ significantly from Eq.~(\ref{freeatom}). In this case, Eq.~(\ref{Es+}) has to be replaced by \cite{Milonni1995} 
\begin{equation}\label{Es++}
	\hat  {\bf E}^+ ({\bf r}, t) = \hat  {\bf E}_{\rm in}^+ ({\bf r}, t) + \boldsymbol{ \Psi}({\bf r})\, \ddot{ \hat{{\sigma}}}_x^{+}  ( \tilde t)\, ,
\end{equation}
where $\hat{{\sigma}}_x^{+}$ describes the positive-frequency component of $\hat \sigma_x$ with $\lim_{\lambda \to 0} \hat{{\sigma}}_x^{+} = \hat \sigma_-$. It can be obtained by expanding $\hat \sigma_x$ in terms of the energy eigenstates of the coupled atom-field system. If $| i \rangle$ are the  eigenstates with eigenvalues $\omega_i > \omega_j$ for $i >j$, we obtain:
$
\hat{{\sigma}}_x^{+}= \sum_{i <j} \sigma_{ij}\, \hat P_{ij}
$,
where $\sigma_{ij}\equiv \langle i | \hat \sigma_x | j \rangle $ and $\hat P_{ij} \equiv | i \rangle \langle j|$.
The second-order time derivative $\ddot{ \hat{{\sigma}}}_x^{+}$ can be directly obtained by using 
$\hat P_{ij}(t) = \hat P_{ij}\, e^{-i \omega_{ji}t}$, where $\omega_{ji} = \omega_j - \omega_i$. It results $\ddot{ \hat{{\sigma}}}_x^{+} = -  \sum_{i <j} \omega^2_{ji}\, \sigma_{ij}\, \hat P_{ij}$. In this case $G^{(2)}({\bf r}, t;{\bf r}, t) \propto  \langle \ddot{ \hat{{\sigma}}}_x^{-} \ddot{ \hat{{\sigma}}}_x^{-} \ddot{ \hat{{\sigma}}}_x^{+} \ddot{ \hat{{\sigma}}}_x^{+}  \rangle $ and, except when $ \lambda\rightarrow 0$, there is no general rule implying $G^{(2)}({\bf r}, t;{\bf r}, t) =0$. 

In order to understand how the USC regime affects the statistics of photons emitted by a single quantum emitter, we consider a generalized quantum Rabi model \cite{Garziano2015a} described by the Hamiltonian ($\hbar =1$):
\be \label{GRabi}
\hat H=  \frac{\omega_a}{2} \hat \sigma_z + \omega_c \hat a^\dag \hat a + \lambda 
(\hat a + \hat a^\dag)\, \hat {{\cal I}_\theta}\,, 
\ee
where $\hat a ^\dag$  and $\hat a$ are the creation and annihilation photon operators for a single-mode cavity, $\omega_c$ is the resonance frequency of the cavity mode, $\hat {\cal I}_\theta = \cos \theta \hat \sigma_x +  \sin \theta \hat \sigma_z$, and 
$ \hat{\boldsymbol{\sigma}} $ are Pauli operators.
 This Hamiltonian includes a longitudinal coupling term ($ \propto \hat \sigma_z  $) which arises from the broken inversion symmetry of the atomic potential energy, and will allow us to treat not only the case of natural atoms, but also artificial atoms realised with superconducting circuits \cite{You2011}. For a flux qubit artificial atom, the flux dependence is encoded in the angle $\theta$: $\sin \theta = \varepsilon / \omega_a$, where $\varepsilon$ is the flux bias and $\omega_a = \sqrt{\Delta^2 + \varepsilon^2}$, with $\Delta$ describing the qubit energy gap in the absence of the flux bias. When the flux bias is zero, $\theta=0$ and Eq.~(\ref{GRabi}) reduces to the standard quantum Rabi Hamiltonian $ \hat H_{\rm R}$. 

In circuit QED experiments, the artificial atoms can be excited by coupling them to a transmission line. Moreover, it is also possible to measure their state by detecting the reflected or emitted electromagnetic field. For example, if a semi-infinite transmission line is terminated with an inductive coupling to an artificial atom, the output voltage $\hat V_{\rm out}$ can be related to the input voltage $\hat V_{\rm in}$ by the following relationship \cite{Girvinlectures}
\be \label{inout}
	\hat V_{\rm out}(x,t) = - \hat V_{\rm in}(x,t) - m\,  \dot{\hat{I}}_a(\tilde t)\, ,
\ee
where $m$ describes the mutual inductance, ${\hat{I}}_a$ is the atom current operator and  $\tilde t = t-x/v$, where $v$ is the speed of light in the transmission line.
For a flux-qubit artificial atom the current operator can be written, in the qubit energy-eigenbasis, as
$\hat I_a = I_a \hat {\cal I}_\theta$. The resulting positive frequency component of the output voltage is
$
	\hat V^+_{\rm out}(x,t) = - \hat V^+_{\rm in}(x,t) - \beta\,  \dot{\hat{{\cal I}}}_\theta^+(\tilde t)
$,
where $\beta =m I_a$. Also in this case, if the interaction of the artificial atom  with the electromagnetic field does not significantly affects  its dynamics, we can approximate $ \dot{\hat{{\cal I}_{\theta}}}^+(\tilde t) \simeq -i \omega_a \hat \sigma_-$ so that $G^{(2)}(x,t,x,t) \propto \langle   \hat \sigma_+  \hat \sigma_+ \hat \sigma_- \hat \sigma_- \rangle =0$.

Our aim is to study the statistics of the photons emitted by a general single two-level system coupled to a photonic resonator for a wide range of light-matter coupling strengths. To this end, we first focus our attention on the quantum Rabi model ($ \theta=0 $) and its generalizations to include the diamagnetic term proportional to the square of the field amplitude \cite{Tufarelli2015,GarciaRipoll2015}. We will then consider the case of artificial atoms where the light-matter longitudinal coupling is present. When the normalized coupling strength $\lambda / \omega_a$ is not sufficiently small, the second order normalized correlation function may depend on the specific operators describing the light emitted by the atom and can be expressed as
\be \label{g2g}
g^{(2)}_{O}(\tau) = \frac{ \langle \hat O^-(t) \hat O^-(t + \tau) \hat O^+(t + \tau)  \hat O^+(t) \rangle}{\langle \hat O^-(t) \hat O^+(t) \rangle \langle \hat O^-(t+ \tau) \hat O^+(t + \tau) \rangle}\, ,
\ee
where $\hat O^\pm$ is a positive or negative frequency component operator. In this Letter, we present calculations for  $\hat O \in \left[\hat \sigma_{x}, \dot{ \hat{{\sigma}}}_x, \ddot{ \hat{{\sigma}}}_x ,\dot{ \hat{{\cal I}}}_{\theta}\right]$. In the limit $\lambda / \omega_a \to 0$, all these operators provide $g^{(2)}_{O}(0)=0$ as a result. 

\begin{figure}[!ht]
	\centering
	\includegraphics[scale=0.68]{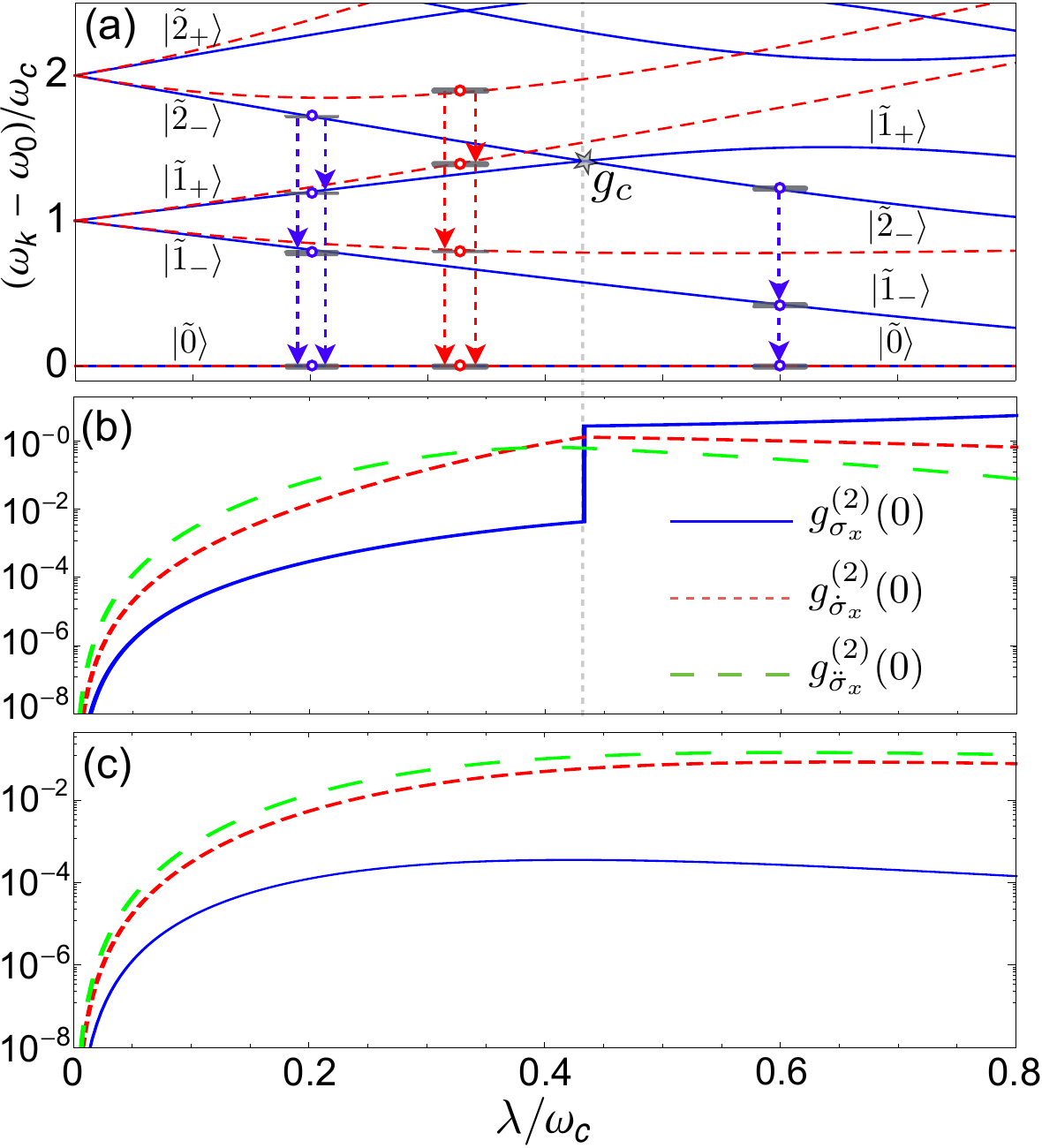}
	\caption{(a) Energy spectra of the quantum Rabi Hamiltonian $\hat H_{\rm R}$ (blue solid curves) and of the Hamiltonian $\hat H_{\rm d}$ including the diamagnetic term  (red dashed curves) as a function of the normalized coupling strength $\lambda / \omega_c$ at zero detuning $ (\omega_c = \omega_a) $. The arrows describe the possible decay channels for the system prepared in the eigenstate $| \tilde 2_- \rangle$ of $\hat H_{\rm R}$ (for  $\lambda / \omega_c < g_c$ and  $\lambda / \omega_c > g_c$) and $\hat H_{\rm d}$ (for $\lambda / \omega_c > g_c$) (b) Normalized second-order correlation functions 
	$g^{(2)}_{O}(0)$ for $ O =\sigma_{x}, \dot{\sigma}_x, \ddot{\sigma}_x$ as a function of $\lambda / \omega_c$, for the system prepared in the initial state $| \tilde 2_- \rangle$ of $\hat H_{\rm R}$ and (c) in the initial state $| \tilde 2_- \rangle$ of $\hat H_{\rm d}$.
	\label{fig:2}}
\end{figure}

Figure~\ref{fig:2}a displays the energy differences between the lowest energy levels and the ground state energy as a function of  the normalized coupling strength $\lambda/ \omega_a$ obtained by numerical diagonalization of the quantum Rabi Hamiltonian $ \hat H_{\rm R} $ (blue solid curves).  The red dashed curves have been obtained diagonalizing the Hamiltonian which includes the diamagnetic term: $\hat H_{\rm d} =	\hat H_{\rm R} + D (\hat a + \hat a^\dag)^2$ with $D = \lambda^2 / \omega_a$. All the results displayed in \figref{fig:2} have been obtained at zero detuning ($\omega_c = \omega_a$).
The ground state is indicated as $|\tilde{0} \rangle$ and the excited  energy states have been labeled as $|\tilde{n}_\pm \rangle$ on the basis of the usual notation for the eigenstates of the Jaynes-Cummings (JC) eigenstates $|{ n_\pm} \rangle$. When the counter-rotating terms in the Hamiltonian go to zero, each state $|\tilde{ n}_\pm \rangle  \to |{ n_\pm} \rangle$ and the two states conserve their parity for all values of $\lambda/ \omega_a$.

We consider the system initially prepared in the state $|\tilde{ 2}_- \rangle$. The arrows in \figref{fig:2}a show the available decay channels. A crossing between the energy levels $\omega_{\tilde{2}_- }$ and $\omega_{\tilde{1}_+ }$ of the quantum Rabi model  can be observed at $\lambda / \omega_a = g_{\rm c} \sim 0.45$. For $\lambda / \omega_a < g_{\rm c}$,  the quantum Rabi model displays two possible decay channels towards the ground state: $|\tilde{2}_-\rangle \to |\tilde{1}_\pm \rangle \to |\tilde{0} \rangle$. Other possible  transitions as, e.g., $|\tilde{1}_+ \rangle \to |\tilde{1}_- \rangle$ or $|\tilde{2}_- \rangle \to |\tilde{0} \rangle$ are forbidden owing to the parity selection rule. For $\lambda / \omega_a  > g_{\rm c}$, only one decay channel is allowed. The resulting zero-delay normalized second order correlation function at $t=0$ can be written as
		
\be \label{g2}
	g^{(2)}_{O}(0) = \frac{|{O}_{\tilde{0},\tilde{1}_{+}}\,  {O}_{\tilde{1}_+,\tilde{2}_-} + 
	{O}_{\tilde{0},\tilde{1}_{-}}\,  {O}_{\tilde{1}_-,\tilde{2}_-}	
		|^2}{(
		|{O}_{\tilde{1}_+,\tilde{2}_-}|^2 + 	|{ O}_{\tilde{1}_-,\tilde{2}_-}|^2)^2}\, ,
\ee
where ${O}_{n,m} \equiv \langle n| \hat O |m \rangle$, when the two decay channels are present, and as
\be\label{g2c}
g^{(2)}_{O}(0) = \frac{|
	{O}_{\tilde{0},\tilde{1}_{-}}\,  {O}_{\tilde{1}_-,\tilde{2}_-}	
	|^2}{|{O}_{\tilde{1}_-,\tilde{2}_-}|^4}\, 
\ee
after the crossing, where there is only one decay channel.

For small coupling strengths ($\lambda / \omega_a \to 0$), where the JC eigenstates are a good approximation, it results ${\sigma}_{0,1_\pm}\,  {\sigma}_{1_\pm,2_-} = \pm 1/2$, and the numerator in Eq.~(\ref{g2}) implies $g^{(2)}_{\sigma_x}(0) \to 0$. Hence the well-know result $g^{(2)}_{\sigma_x}(0) = 0$ can be interpreted as the result of complete destructive interference of the two possible paths determining the two terms in the numerator of Eq.~(\ref{g2}).
Figure~\ref{fig:2}b displays the three $g^{(2)}_{O}(0)$ with $ O=\sigma_x,\dot{\sigma}_x,\ddot{\sigma}_x $ for the quantum Rabi model as a function of the normalized coupling strength $\lambda / \omega_a$. As expected, the three curves start from zero for $\lambda / \omega_a \to 0$ and increase for increasing values of the coupling strength. We notice that $g^{(2)}_{\sigma_x}(0)$ (blue solid curve) remains below $10^{-3}$
for $\lambda / \omega_a < g_c$, since the corrections due to the counter-rotating terms are not able to affect the products ${\sigma}_{\tilde 0,\tilde 1_\pm}\,  {\sigma}_{\tilde 1_\pm,\tilde 2_-}$ so that the numerator of Eq.~(\ref{g2}) remains very small. When $\lambda / \omega_a = g_c$, a sharp transition occurs. As shown in \figref{fig:2}a, for $\lambda / \omega_a > g_c$ only one decay channel is allowed and  no cancellation effects are possible in the numerator of $g^{(2)}_{ \sigma_x}(0)$ (see Eq.(\ref{g2c})) which jumps to $\sim 3$. 
The situation is different for $g^{(2)}_{\dot{ \sigma}_x}(0)$ and  $g^{(2)}_{\ddot{ \sigma}_x}(0)$. For example, if we consider $g^{(2)}_{\dot{ \sigma}_x}(0)$ we observe that the two terms in the  numerator in Eq.~(\ref{g2}) for $\lambda / \omega_a < g_c$ have the form $ \omega_{\tilde 1_\pm,\tilde 0} \omega_{\tilde 2_-,\tilde 1_\pm} {\sigma}_{\tilde 0,\tilde 1_\pm}\,  {\sigma}_{\tilde 1_\pm,\tilde 2_-}$.
 Even if  ${\sigma}_{\tilde 0,\tilde 1_\pm}\,  {\sigma}_{\tilde 1_\pm,\tilde 2_-} \simeq \pm 1/2$, for increasing values of the coupling strength the transition frequency $\omega_{\tilde  2_-, \tilde 1_+}$ decreases significantly, lowering one of these two terms in the numerator so that $g^{(2)}_{\dot{ \sigma}_x}(0)$ can differ significantly from zero even before $\lambda / \omega_a = g_c$.
Figure~\ref{fig:2}c displays the three $g^{(2)}_{O}(0)$ for the system described by $\hat H_{\rm d}$ as a function of the normalized coupling strength $\lambda / \omega_a$ under the same conditions used to derive the results in \figref{fig:2}b. As shown in \figref{fig:2}a, in this case no level crossings occur and two decay channels are always present. The normalized correlation functions are described by  Eq.~(\ref{g2}). Owing to the destructive interference, $g^{(2)}_{ \sigma_x}(0)$ remains very small (below $10^{-4}$) even at larger coupling strengths. On the contrary, owing to the differences between the transition frequencies of the two available decay paths, $g^{(2)}_{\dot{ \sigma}_x}(0)$ and  $g^{(2)}_{\ddot{ \sigma}_x}(0)$ display experimentally detectable deviations from the standard (weak coupling) result $g^{(2)}_{O}(0)= 0$.
The system can be experimentally prepared   by exciting the qubit or the cavity with a sequence of two resonant $\pi$-pulses, determining the sequential  transitions $|\tilde{0} \rangle \to |\tilde{1}_- \rangle \to |\tilde{2}_- \rangle$, or by two-photon excitation $|\tilde{0} \rangle  \to |\tilde{2}_- \rangle$.  
In this latter case, the resulting state will be the superposition $| \psi (0) \rangle =\sqrt{1 - |\alpha|^2} |\tilde{0} \rangle +  \alpha|\tilde{2}_- \rangle$ with $\alpha \ll 1$. Using this as initial state, the obtained $g^{(2)}_{O}(0)$ can be significantly higher, reproducing the curves displayed in \figref{fig:2}b and \ref{fig:2}c divided, however, by the factor $|\alpha|^2$. 
\begin{figure}[!ht]
	\centering
	\includegraphics[scale=0.5]{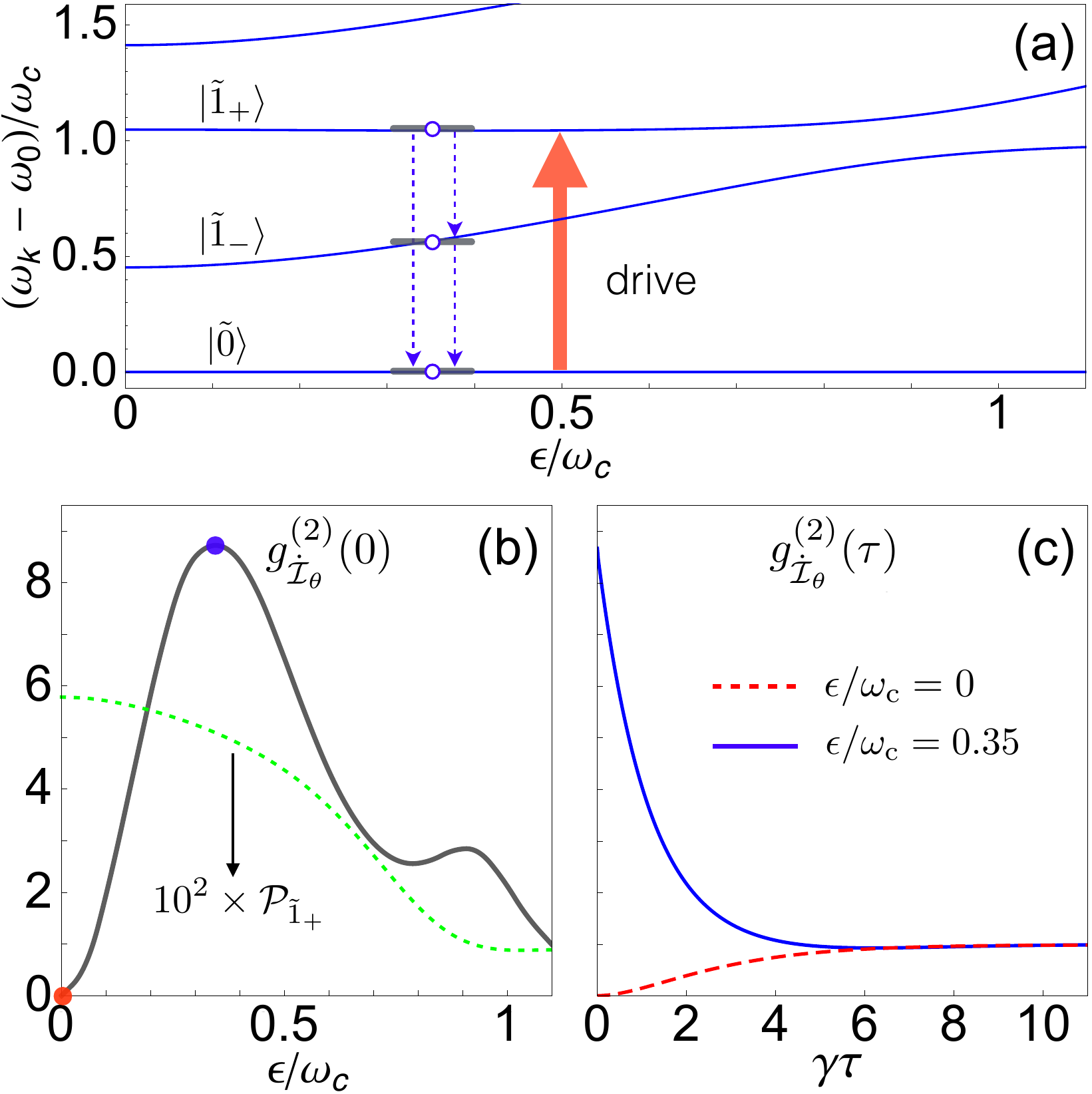}
	\caption{(a) Energy spectrum of the generalized  Rabi Hamiltonian $\hat H$ (blue solid curves) as a function of the normalized flux offset $\varepsilon/\omega_c$. The dashed blue arrows describe the possible decay channels for the system prepared in the eigenstate $| \tilde 1_+ \rangle$, while the solid red arrow describes the system excitation by an applied drive. (b) Qubit normalized second-order correlation function $g^{(2)}_{\dot{\cal  I}_\theta}(0)$ as a function of  $\varepsilon/\omega_c$ (grey solid curve) for the system driven towards the $| \tilde 1_+ \rangle$ state by applying to the resonator a continous-wave optical drive of amplitude $A/ \gamma = 0.25$. The steady-state population $ \mathcal{P}_{\tilde{1}_+} $ of the $| \tilde 1_+ \rangle$ state is also dispayed (green dotted curve). (c) Qubit steady-state time-delayed  normalized second-order correlation function $g^{(2)}_{\dot{\cal I}_\theta}(\tau)$ obtained for $\varepsilon/\omega_{c} = 0$ (red dashed curve) and for  $\varepsilon/\omega_{c} = 0.35$ (blue solid curve). The decay rates for the qubit and the resonator are $\gamma = \kappa = 5 \times 10^{-4}\,  \omega_c$.
		\label{fig:3}}
\end{figure}	
Figure~\ref{fig:3}  shows results in the case for which a longitudinal coupling between the qubit and the resonator field is present ($ \theta\neq0 $). We use a zero-bias qubit gap $\Delta / \omega_c = 0.5$ GHz and a qubit-resonator coupling rate $\lambda / \omega_c = 0.2 $. Panel 3a displays the energy spectrum for the lowest energy levels of $\hat H$ as a function of the normalized flux bias $\varepsilon/\omega_c$.  The arrows describe the possible decay channels for the system prepared in the eigenstate $| \tilde 1_+ \rangle$. For $\varepsilon = 0$, the parity selection rule implies that only the one-photon transition $| \tilde 1_+ \rangle \to | \tilde 0 \rangle$ and as a consequence $g^{(2)}_{O}(0) =0$ is expected. For $\varepsilon \neq 0$, the parity selection rule is broken and all the available downward spontaneous transitions are allowed.
In order to present results that can be experimentally studied more easily, we calculate the steady-state  normalized correlation function $g^{(2)}_{\dot{\cal  I}_\theta}(0)$ for the field emitted by the qubit after continuous-wave pumping of the resonator with a drive resonant with the transition $| \tilde 0 \rangle \leftrightarrow | \tilde 1_+ \rangle$.  The system Hamiltonian including the drive is $\hat{H} + A \cos{(\omega_d t)} (\hat a + \hat a^\dag)$ with $\omega_d = \omega_{\tilde 1_+, 0}$. The influence of the cavity-field damping and atomic decay on the process are taken into account by using the master-equation
approach in the dressed picture \cite{Beaudoin2011, Ridolfo2012}. We consider the system interacting with zero-temperature baths. The master equation is obtained by using the Born-Markov approximation without the post-trace RWA \cite{Law2015}. We use  for the decay rates of the qubit ($\gamma$) and the cavity ($\kappa$): $\gamma = \kappa = 5 \times 10^{-4}\,  \omega_c$. Moreover, we use an excitation amplitude $A/ \gamma = 0.25$  able to provide a steady-state population for the state $|\tilde{1}_+ \rangle$ ranging between 1\% and 6\% depending on the value of $\varepsilon$ (see \figref{fig:3}b).
According to Eq.~(\ref{inout}), the measured second-order correlation function can be obtained from Eq.~(\ref{g2g}) by using $\hat O = \dot{\hat{{\cal I}_\theta}}$.
%The resulting  $g^{(2)}_{\dot{\cal  I}}(0)$ calculated at steady state ($\gamma t \gg 1$) is zero for $\varepsilon = 0$ and reaches its maximum $[g^{(2)}_{\dot{\cal  I}}(0)] \sim 4.8$ at $\varepsilon/\omega_{c} \sim 0.34$.
Figure~\ref{fig:3}b shows the steady-state, zero-delay qubit normalized correlation function  as a function of the flux bias.
Also in this case, we find that it is significantly different from zero, reaching its maximum ($\sim 8.7$) at $\varepsilon/\omega_{c} \sim 0.35$. The shape of the curve in \figref{fig:3}b and the position of its maximum depends on the dependence on $\varepsilon$ of the matrix elements entering Eq.~(\ref{g2g}). 
Figure~\ref{fig:3}c displays the qubit steady-state time-delayed  normalized second-order correlation function $g^{(2)}_{\dot{\cal I}_\theta}(\tau)$ obtained for $\varepsilon/\omega_{c} = 0$ (red-dashed curve) and for  $\varepsilon/\omega_{c} = 0.35$ (where  $g^{(2)}_{\dot{\cal  I}_\theta}(0)$ is maximum) (blue continuous curve). For $\varepsilon/\omega_{c} = 0$, the expected antibunching effect: $g^{(2)}_{\dot{\cal I}_\theta}(\tau) > g^{(2)}_{\dot{\cal I}_\theta}(0)$
can be observed. On the contrary, the blue continuous curve shows a photon bunching effect: $g^{(2)}_{\dot{\cal I}_\theta}(\tau) < g^{(2)}_{\dot{\cal I}_\theta}(0)$.
As expected, both the two curves tend to 1 for $\tau \to \infty$, indicating the loss of correlation between the emitted photons.

We have thus proved that a single two-level system, physically instantiated in a real or an artificial atom, can emit bunched light. The effect here described provides clear evidence of how an atom can lose its identity when ultrastrongly interacting with a photonic resonator. %when ultrastrongly coupled to a photonic resonator. As the observation of bunched light has been a final proof of the quantization of the electromagnetic field, the effect we describe demonstrates how in the USC regime our intuitive understanding of light-matter interaction in terms of emission and absorption processes fails.
%{\color{red} ALTERNATIVA!
%We have thus proved that a single two-level system, physically instantiated in a real or an artificial atom, can emit bunched light when ultrastrongly coupled to a photonic resonator. This effect can be interpreted as clear evidence of how in this light-matter coupling regime, where a deep hybridization occurs between the cavity field and the atom, the latter can lose its identity (OPPURE its own individual features).}
%========== Bibliography =============
\bibliography{Garzbib2}

\end{document}